\documentclass[12pt,oneside,reqno]{amsart}

\newcommand{\Tr}{\texttt{Tr}}

\usepackage{amssymb}
\usepackage{bm}
\usepackage{footnote}

\begin{document}

\title{Hamiltonian Brownian motion \\ %
in Gaussian thermally fluctuating potential. \\
I. Exact Langevin equations, \\ %
invalidity of Marcovian approximation,  \\
common bottleneck of dynamic noise theories, \\
and diffusivity/mobility 1/f noise}

\author{Yu. E. Kuzovlev}

\address{Donetsk Institute for Physics and Technology of NASU,
ul.\,R.\,Luxemburg 72, Donetsk 83114, Ukraine}
\email{\, kuzovlev@fti.dn.ua}





\begin{abstract}
Dynamical random walk of classical particle in  %
thermodynamically equilibrium fluctuating medium, - %
Gaussian random potential field, - is considered %
in the framework of explicit stochastic representation of deterministic %
interactions. We discuss corresponding formally exact Langevin equations %
for the particle's trajectory and show that %
Marcovian kinetic equation approximation to them %
is inadequate, - even (and especially) in case of %
spatially-temporally short-correlated field, - %
since ignores such actual effects of exponential %
instability of the trajectory
(in respect to small perturbations) as  %
scaleless low-frequency diffusivity/mobility fluctuations %
(and other excess degrees of randomness)
reflected by third-, fourth- and higher-order %
long-range irreducible statistical correlations. %
We try to catch the latter, - squeezing through typical %
theoretical narrow bottleneck, - with the help of an exact relationship %
between the instability and diffusivity statistical characteristics, %
along with standard analytical d approximations. The result is quasi-static %
diffusivity fluctuations which generally are comparable with mean value of %
diffusivity and disappear in the limit of infinitely large medium's %
correlation length or infinitely small correlation time %
only, in agreement with the previously suggested %
theorem on fundamental 1/f noise.

\,\,\,

\end{abstract}


\maketitle

\baselineskip 18 pt

\markboth{}{}



\section{Introduction}

In this paper we shall touch classical analogue of interesting %
quantum statistical-mechanical problem already touched in \cite{j,o}. %
Namely, statistics of dynamical ``random walk'', %
or ``Brownian motion'', of (microscopic) particle %
interacting with thermodynamically equilibrium scalar %
boson field. For instance, with (harmonic) phonon field %
(crystal lattice or other medium oscillations). %

In \cite{j,o}, basing on the widely known Hamiltonian model %
of such interaction (``polaron model''), we obtained %
an exact system of shortened evolution equations for probabilitity %
distribution of the ``Brownian particle'' (BP) and its %
statistical correlations with the ``phonon field'' %
(``boson thermostat''), and argued that exact solutions %
to these equations includes 1/f\,-type (scaleless) %
low-frequency fluctuations of BP's diffusivity and thus mobility. %
But direct formulations of these solutions, or %
at least good enough approximation to them (without loss %
of the diffusivity/mobility 1/f\,-noise), still are absent. %
Therefore, for further progress in statistical-mechanical %
theory of fundamental 1/f-noise, it may be useful to consider %
classical variant of the mentioned Hamiltonian model. %

At that, we can avoid any detailing of ``phonon'' (thermostat) %
part of full Hamiltonian, if instead exploit the ``stochastic %
representation'' (SR) of dynamical (deterministic) interactions, %
for the first time suggested and tested in \cite{sr1,sr2} and later %
generalized, developed and applied in \cite{sr3,sr4,sr5,sr6,sr7}. %
Thus our consideration will be %
at once additional probing of old and search of new SR possibilities. %

\section{Principles of the stochastic representation}

Firstly we have to recall main SR statements \cite{sr1,sr2,sr3,sr5}. %
Let a Hamiltonian system consists of a ``Dynamical subsystem''
under our interest, ``D'', and some its environment, or %
``thermal Bath'' (thermostat), ``B'', and full Hamiltonian of  %
``D+B'' has bilinear form
\begin{equation}
H\,=\,H_d + H_b+H_{int}\,\,,\,\,\, %
H_{int}\,=\sum_n \,D_n\, B_n\,\,, \, \label{h}
\end{equation}
where operators (or phase functions, in
classical mechanics) $\,H_d\,,\,D_n\,$ and $\,H_b\,,\,B_n\,$ %
are defined in Hilbert spaces (or phase spaces) of ``D'' and
``B'', respectively. %
Besides, let initially, somewhen in the past, full density matrix %
(probability distribution function) of ``D+B''
$\,\rho(t)\,$ had factored form:\, %
$\,\rho^{(in)}=\rho(t\rightarrow -\infty)= %
\,\rho_d^{(in)}\,\rho_b^{(in)}\,$. %
Then marginal density matrix (DM) of ``D``, %
$\,\rho_d(t)= \Tr_b\, \rho(t)\,$, can be expressed %
as average value,
\begin{eqnarray}
\rho_d(t)\,=\,\langle\, \widetilde{\rho}(t)\,\rangle\,\,,
\,\label{av}
\end{eqnarray}
 of randomly varying DM $\,\widetilde{\rho}(t)\,$ satisfying stochastic %
von Neumann (or Liouville) evolution equation
\begin{eqnarray}
\frac {d\widetilde{\rho}(t)}{dt}\,=\, %
\frac i\hbar\,\left[\,\widetilde{\rho}(t)\,,\,  %
H_d +\sum_n\,x_n(t)\,D_n\,\right]\, +\, %
\sum_n\,y_n(t)\,D_n\circ \widetilde{\rho}(t)\,\,, \,
\label{ee}
\end{eqnarray}
where\, $\,\circ\,$\, denotes symmetrized (Jordan) product, %
$\,A\circ B=(AB+BA)/2\,$, %
and $\,x_n(t)\,$ and $\,y_n(t)\,$ are random processes. %

At that, all statistical characteristics of %
$\,x_n(t)\,$ and $\,y_n(t)\,$ are unambiguously determined %
by internal dynamical properties of ``B'', along %
with its initial DM, $\,\rho_b^{(in)}\,$. Corresponfing formulae %
can be found in \cite{sr1,sr2,sr3,sr4,sr5} %
(for most general variants of SR, %
including non-Hamiltonian dynamics, see \cite{sr3}). %
In particular, if ``B'' is a set (continuum) %
of harmonic oscillators (wave modes), %
while $\,\rho_b^{(in)}\,$ has canonical Gibbs form, with some %
temperature $\,T\,$, then $\,x_n(t)\,$ and $\,y_n(t)\,$ %
are stationary Gaussian random processes representing %
thermodynamically equilibrium ``Gaussian thermostat''. %

Below, we confine ourselves by this case, but generalize it %
to continuous index $\,n\,$\,:
\begin{eqnarray}
\sum_n\,D_n\,B_n\, \Rightarrow\, %
\int D(r)\,B(r)\,dr\,\,, \, \label{csr}\\
\sum_n\,x_n(t)\,D_n\, \Rightarrow\, %
\int x(t,r)\,D(r)\,dr\,\,, \,\,\,\,\, %
\sum_n\,y_n(t)\,D_n\, \Rightarrow\, %
\int y(t,r)\,D(r)\,dr\,\,, \, \nonumber %
\end{eqnarray}
where $\,r\,$ marks points of a $\,d\,$-dimensional %
space, and $\,dr=d^dr\,$.

\section{Random fields of equilibrium thermostat}

It is important to remind, firstly, that %
$\,x_n(t)\,$ and hence $\,x(t,r)\,$ represent direct dynamical %
perturbation of ``D'' by ``B'', i.e. thermostat noise, %
while $\,y_n(t)\,$ and $\,y(t,r)\,$  inverse %
perturbation of ``B'' by ``D'' and related feedback action %
of ``B'' onto ``D'', in particular, ``friction'' %
or ``viscosity'', etc., i.e. thermostat induced dissipation. %
Therefore $\,y_n(t)\,$ or $\,y(t,r)\,$ are peculiar (``ghost'') %
random variables:\, all their self-correlation are zeros, %
e.g.\, $\,\langle y(t_1,r_1)\,y(t_2,r_2)\rangle =0\,$,\, althouh %
their cross-correlations with $\,x_n(t)\,$ or $\,x(t,r)\,$ %
can differ from zero.

Secondly, the generalized fluctuation-dissipation %
relations (FDR\,,\, see e.g. \cite{bkn,p1106,p1108, %
p0501,fds1,fds2,fds3,ph1,ph2,ufn} and references therein), %
resulting from fundamental properties of Hamiltonian dynamics, %
imply definite mutual correspondence of the ``x-y'' %
cross-correlators and ``x-x'' self-correlators . %
In particular, for equilibrium and spatially homogeneous %
thermostat, according to standard recipes from %
\cite{sr1,sr2,sr3,sr4}, we can write
\begin{eqnarray}
K_{xx}(\tau,r_1-r_2)\,\equiv\, %
\langle \,x(t+\tau,r_1)\,x(t,r_2)\,\rangle\,=\, %
\int_{-\infty}^\infty \cos\,\omega\tau \, %
\,S(\omega,r_1-r_2)\, \frac {d\omega}{2\pi}\,\,, \, \nonumber\\
K_{xy}(\tau,r_1-r_2)\,\equiv\, %
\langle \,x(t+\tau,r_1)\,y(t,r_2)\,\rangle\,=\, %
\label{cf}\\ \,=\, %
-\,\theta(\tau)\, \frac 2\hbar \int_{-\infty}^\infty %
\sin\,\omega\tau \, \,\tanh\,\frac{\hbar\omega}{2T}\,\, %
S(\omega,r_1-r_2)\, \frac {d\omega}{2\pi}\,\,, \, \nonumber
\end{eqnarray}
where $\,\theta(\tau)\,$ is Heaviside step function, %
thus expressing both ``x-x'' and ``x-y'' correlators %
through one and the same %
spectral function, %
$\,S(\omega,r_1,r_2)\geq 0\,$\,\,\footnote{\, %
Notice that in \cite{sr1,sr2} the minus sign %
in (\ref{cf}) was absent, but this misprint %
had not penetrated to results of SR applications.}\,.
Importantly, in accordance with the causality principle, %
$\,y(t,r)\,$ is only correlated %
with later $\,x(t^\prime >t,r^\prime)\,$. %
In the classical limit, the FDR (\ref{cf}) reduces to
\begin{eqnarray}
K_{xy}(\tau,r)\,=\, \frac {\theta(\tau)}{T}\, %
\frac{\partial}{\partial \tau}\, K_{xx}(\tau,r)\,\, %
\label{ccf}
\end{eqnarray}
In case of harmonic thermostat, these two correlators %
completely determine (Gaussian) statistics of %
$\,x(t,r)\,$ and $\,y(t,r)\,$.

\section{Particle in thermal random field}

To consider ``Brownian particle'' (BP) in a thermally fluctuating %
media, let us model the latter with potential field %
and choose in (\ref{h})-(\ref{csr})
\begin{eqnarray}
H_d\,\Rightarrow\, \frac {P^2}{2m}\,\,, \,\,\,\,\, %
D_n\, \Rightarrow\, D(r)\, =\,\delta(r-R)\,\,, \,
\label{hh}
\end{eqnarray}
with $\,P\,$ and $\,R\,$ being (operators of) BP's momentum %
and coordinate and $\,B(r)\,$ (operator of) %
the fluctuating potential. %
Then, in the Wigner representation, %
the ``stochastic von Neumann (quantum Liouville) %
equation'' (\ref{ee}) takes form
\begin{eqnarray}
\frac {\partial \widetilde{\rho}}{\partial t}\,=\, %
- \frac Pm\,\nabla_{\!\! R} \,\widetilde{\rho} \,+\, %
\frac i\hbar\,\left[\,x\left(t,R - %
\frac {i\hbar}{2}\,\nabla_{\!\! P} \right) - %
x\left(t,R + \frac {i\hbar}{2}\,\nabla_{\!\! P} \right) %
\right]\,\widetilde{\rho}\,+\, %
\nonumber\\ \, +\, %
\frac 12 \left[\,y\left(t,R - %
\frac {i\hbar}{2}\,\nabla_{\!\! P} \right) + %
y\left(t,R + \frac {i\hbar}{2}\,\nabla_{\!\! P} \right) %
\right]\,\widetilde{\rho}\,\, \label{qee}
\end{eqnarray}
with $\,\nabla_{\!\! R}\,$ and $\,\nabla_{\!\! P}\,$ denoting %
derivatives (gradients). In the classical limit, clearly, %
it turns to ``stochastic Liouville equation'' (SLE)
\begin{eqnarray}
\frac {\partial \widetilde{\rho}}{\partial t}\,=\, %
- \frac Pm\,\nabla_{\!\! R}\, \widetilde{\rho} \,+\, %
\nabla x(t,R)\, \nabla_{\!\! P} \,\widetilde{\rho}\,+\, %
y(t,R) \,\widetilde{\rho}\,\,, \, \label{cee}
\end{eqnarray}
where $\,\nabla x(t,R)= \nabla_{\!\! R}\, x(t,R)\,$. %

\section{Langevin equations}

Natural solution to the SLE (\ref{cee}) is
\begin{eqnarray}
\widetilde{\rho}(t,R,P)\,=\, %
\delta(R-R(t))\, \delta(P-P(t))\, %
\exp\,[\,\int_{t>t^\prime} y(t^\prime,R(t^\prime))\, %
dt^\prime\,]\,\,, \, \label{s}
\end{eqnarray}
where $\,R(t)\,$ and $\,P(t)\,$ are random processes %
obeing stochastic ODE
\begin{eqnarray}
\frac {dR(t)}{dt} = \frac {P(t)}m\,\,, \,\,\,\,\, %
\frac {dP(t)}{dt} = -\nabla x(t,R(t))\,\,, \,
\label{se}
\end{eqnarray}
with some (may be random) initial conditions %
in the past. %

These equations do not display dissipative feedback %
action of the media (i.e. BP's self-action through media), %
which however is completely involved by field $\,y(t,r)\,$ %
and ecomes apparent after averaging expression (\ref{s}) %
to get (\ref{av}). %

At the same time, %
classical SR allows \cite{sr5} %
to perform in (\ref{s}) separate %
averaging over ``x-y'' cross-correlations, %
in such way removing $\,y(t,r)\,$ from (\ref{s}) %
and dispaying its effect in (\ref{se}). %
As the result, stochastic equations (\ref{se}) %
transform to what can be called ``Langevin equations'' (LE). %
A general recipe to construct LE is accumulated by %
Eqs.39-42 in \cite{sr5}. In case of Gaussian thermostat, %
it strongly simplifies (see Eqs.44 %
and following example in \cite{sr5}), %
and in application to our present system, %
as defined by Eqs.\ref{hh}, yields %
\begin{eqnarray}
\frac {dR(t)}{dt} \,=\, \frac {P(t)}m\,\,, \, \nonumber\\ %
\frac {dP(t)}{dt} \,=\, f(t,R(t))\,- %
\int_{t>t^\prime} %
K(t-t^\prime,R(t)-R(t^\prime))\, dt^\prime\,\, %
\, \label{le}
\end{eqnarray}
Here, the integral represents medium's feedback response,, %
introducing friction and dissipation, %
its (vector-valued) kernel $\,K(t,r)\,$ is expressed by\,
\begin{eqnarray}
K(t,r)\,=\, \nabla_{\!\! r} K_{xy}(t,r)\,=\,  %
\frac {1}{T} \int\int \sin\,kr\,\, %
\sin\,\omega\tau\,\, k\omega\, S(\omega,|k|)\, %
\frac {d\omega}{2\pi}\,\frac {d^dk}{(2\pi)^d}\, \,
\,\,, \, \label{k}
\end{eqnarray}
and \,$\,f(t,r))= -\nabla x(t,r)\,$\, is feedback-free ``seed'' %
medium's random force field possessing %
Gaussian statistics with zero average value, %
\begin{eqnarray}
\langle\, f(t,r)\,\rangle \,=\,0\,\,, \, \label{z}
\end{eqnarray}
and correlation function
\begin{eqnarray}
\langle\,f(t_1,r_1)*f(t_2,r_2)\,\rangle\,=\, %
\nabla_{r_1} *\nabla_{r_2}\, %
K_{xx}(t_1-t_2,r_1-r_2)\,\, \,\label{fc}
\end{eqnarray}
($\,\,*\,\,$ denotes tensor product of vectors). %
Now the random distribution function (DF) to be %
inserted to (\ref{av}) is, instead of (\ref{s}), merely %
$\,\widetilde{\rho}(t)=$ $\delta(R-R(t))\, \delta(P-P(t))\,$.

Of course, BP's state  %
$\,\{R(t),P(t)\}\,$ is  densely correlated %
with the force $\,f(t,R(t))\,$, %
therefore Gaussianity of the field $\,f(t,r)\,$ in itself %
does not mean Gaussianity of $\,f(t,R(t))\,$. %

\section{Quasi-quantum formulation}

Excluding from theory, - in the spirit of quantum mechanics, - %
BP's momentum, and considering  %
BP's coordinate marginal distribution only,
\[
\widetilde{W}(t,R)\,=\,\int \widetilde{\rho} %
(t,R,P)\,d^dP\,\,, \,
\]
one can derive for it, instead of (\ref{cee}), equations
\begin{eqnarray}
\frac {\partial \widetilde{W}(t,R)}{\partial t}\,=\, %
- \nabla [V(t,R)\,\widetilde{W}(t,R)]\,+\, %
y(t,R)\,\widetilde{W}(t,R)\,\,, %
\, \label{qce}\\
V(t,r)\,=\, -\, \frac 1m\, \nabla %
A(t,r)\,\,, \, \nonumber
\end{eqnarray}
where $\,V(t,r)\,$ is random velocity field generated by %
scalar ``action'' field $\,A(t,r)\,$ which %
satisfies nonlinear stochastic %
Hamilton-Jacobi equation:
\begin{eqnarray}
\frac {\partial A(t,r)}{\partial t}\,=\, %
\frac 1{2m}\,(\nabla A(t,r))^2 \,+\,x(t,r)\, %
\, \, \label{hj}
\end{eqnarray}
Then, again performing partial averaging in respect to ``x-y'' %
cross-correlations, we can transform this into
\begin{eqnarray}
\frac {\partial \widetilde{W}}{\partial t}\,=\, %
- \nabla [V\,\widetilde{W}]\,\,,  %
\,\,\,\,\, V\,=\, -\, \frac 1m\, \nabla %
A\,\,, \,\,\,\, \label{qce_}\\
\frac {\partial A}{\partial t}\,=\, %
\frac {(\nabla A)^2}{2m}\,+\,x(t,r)\,+\, %
\int_{t>t^\prime} \int K_{xy}(t-t^\prime,r-r^\prime)\, %
\widetilde{W}(t^\prime,r^\prime)\,d^dr^\prime\,dt^\prime\, %
\, \, \,\,\, \label{hj_}
\end{eqnarray}
Obviously, the latter stochastic PDE envelopes Eqs.\ref{le} %
and serves as a kind of LE. %

\section{Concretization of problem}

It is seems reasonable to assume, first, %
that our fluctuating media is statistically %
isotropic, therefore functions from (\ref{cf}) are %
spherically symmetric functions of coordinate differences, %
and we can write
\begin{eqnarray}
S(\omega,r)\,=\,  %
\int \cos\,kr\,\, S(\omega,|k|)\, %
\frac {d^dk}{(2\pi)^d}\,\, \, \nonumber
\end{eqnarray}
(with, clearly, $\,S(-\omega,\kappa)=S(\omega,\kappa)\,$). %

Second, the force field is not too singular, %
so that BP's momentum diffusivity is finite:
\begin{eqnarray}
\int_0^\infty \langle\,f(t,R(t)) * f(t-\tau,R(t-\tau)) %
\,\rangle\,d\tau\, \sim\, %
\nonumber\\ \sim\, %
\int_0^\infty [\,-\nabla *\nabla\, %
K_{xx}(\tau,R(t)-R(t-\tau))\,]\, %
d\tau\, \equiv\, D^{(P)}\, %
\, \neq\,\infty\,\,\, \label{c0}
\end{eqnarray}
This condition holds only if
\begin{eqnarray}
\int k^2\,S(Vk,|k|)\, d^dk\, <\,\,\infty\, \label{c}
\end{eqnarray}
at any velocity $\,V\,$. %
Then BP's momentum and velocity, %
$\,V(t)\equiv dR(t)/dt =P(t)/m\,$, %
behave as continuous random processes. %

Third, at any $\,k\,$,
\begin{eqnarray}
\int S(\omega,|k|)\, d\omega\, <\,\,\infty\,\, \, \label{c1}
\end{eqnarray}
This means that $\,f(t,r)\,$ %
possesses nonzero correlation time, $\,\tau_c\,$, %
i.e. is not a ``white noise'' in time. %
Evidently, otherwise space-time variations of $\,f(t,r)\,$ %
would be able to propagate with infinite velocity. %
which would be rather nonphysical behavior. %

Fourth, the spectral function $\,S(\omega,|k|)\,$ %
is such that
\begin{eqnarray}
\lim_{\tau \rightarrow 0}\, %
\,\nabla K_{xx}(\tau,V\tau))\,\, %
=\,0\,\,\, \nonumber 
\end{eqnarray}
Then, using FDR (\ref{ccf}) and %
calculating the feedback integral in Eq.\ref{le} by parts, %
we can transform Eqs.\ref{le} into
\begin{eqnarray}
\frac {dR(t)}{dt} \,=\,V(t)\,\,, \, \nonumber\\
\frac {dV(t)}{dt} \,=\, f(t,R(t))/m\, %
- \int_{t>t^\prime} %
G(t-t^\prime,R(t)-R(t^\prime))\,V(t^\prime)\, dt^\prime\,\,, %
\, \label{le2}
\end{eqnarray}
where matrix response function
\begin{eqnarray}
G(\tau,r) \,=\, %
\langle\,f(\tau,r) * f(0,0) %
\,\rangle/mT \,=\, %
- \nabla *\nabla K_{xx}(\tau,r)/mT\, %
\,=\,  \nonumber\\ =\, %
\frac {1}{mT} \int\int \cos\,kr\,\, %
\cos\,\omega\tau\,\, k*k\,S(\omega,|k|)\, %
\frac {d\omega}{2\pi}\, dk\, \,\label{g}
\end{eqnarray}
determines BP'viscous friction %
(here and below $\,dk\equiv d^dk/(2\pi)^d\,$). %

It is interesting task to reveal, under enumerated %
conditions, long-time statistical %
properties of BP's displacement, %
e.g. $\,\Delta R(t) =R(t)-R(0)\,$.

\section{Likely  conjectures %
and Marcovian approximation}

\subsection{Time-local friction approximation}

\,Notice that because of condition (\ref{c1}) the friction %
always is more (may be highly) or less (may be slightly) %
nonlinear in respect to BP's velocity. %

If characteristic velocity's relaxation time, %
\begin{eqnarray}
\tau_v\,\sim\, mT/D^{(P)}\, \, \, \nonumber %
\end{eqnarray}
(with $\,D^{(P)}\,$'s diagonal in mind), %
is large enough in comparison with that of %
$\,G(\tau,V\tau)\,$ (i.e. the force's correlation 
time $\,\tau_c\,$), %
then one can visualize the non-linearity by %
``time-local approximation'' of LE (\ref{le2}),
\begin{eqnarray}
dR(t)/dt\,=\,V(t)\,\,, \, \, \nonumber\\
dV(t)/dt \,=\, f(t,R(t))/m\, %
- g(V(t))\,V(t)\,\,, \, \label{ale}
\end{eqnarray}
with time-local relaxation rate
\begin{eqnarray}
g(V)\,\equiv\, \int_0^\infty G(\tau,V\tau))\,d\tau\,=\, %
\frac 1{2mT} \int k*k\,S(Vk,|k|)\, dk\, %
\sim\, \frac 1{\tau_v} \,\,. \,  \label{ag}
\end{eqnarray}

\subsection{Marcovian approximation}

\,Simultaneously, the above approximation %
pushes to treat $\,f(t,R(t))\,$ %
like delta-correlated (``white'') noise. %
In order to reasonably determine its characteristics, %
we have to consider evolution equation for %
the random DF %
$\,\widetilde{\rho}(t)=\widetilde{\rho}(t,R,V)\,$, %
corresponding to Eqs.\ref{ale}, i.e. SLE %
\begin{eqnarray}
\partial \widetilde{\rho}/\partial t\,=\, %
\{- V\nabla_{\!\! R}\,+\, %
\nabla_{\!\! V}\,[\,g(V)\,V\,-\,f(t,R)\,]\}\, %
\widetilde{\rho}\,\,, \, \label{aee}
\end{eqnarray}
and coarsen it into approximate kinetic (Fokker-Planck) equation %
for actual BP's DF $\,\rho_d(t)=\rho_d(t,R,V)\,$ (\ref{av}). %
Quite standard manipulations yield
\begin{eqnarray}
\partial \rho_d(t)/\partial t\,=\, %
[- V\nabla_{\!\! R}\,+\,\nabla_{\!\! V}\,g(V)\,V\,]\,\rho_d(t)\, %
- \nabla_{\!\! V} \,m^{-1}\,\langle\,f(t,R)\,\widetilde{\rho}(t)\, %
\rangle\,\,, \,\nonumber\\
-m^{-1}\,\langle\,f(t,R)\,\widetilde{\rho}(t)\, %
\rangle\, \approx \, %
\nonumber\\ \approx\, %
\frac 1{m^2} %
\int_0^\infty \langle f(t,R)\, %
e^{-\tau V \nabla_{\!\! R}}\, f(t-\tau,R)\, %
\nabla_{\!\! V}\, %
\widetilde{\rho}(t-\tau)\,\rangle\, %
d\tau\, \approx\, \label{e1}\\ \,\approx\, %
\left[\,\frac 1{m^2} %
\int_0^\infty \langle f(t,R)\, %
e^{-\tau V \nabla_{\!\! R}}\, f(t-\tau,R)\,\rangle\, %
\nabla_{\!\! V}\, e^{\,\tau V \nabla_{\!\! R}}\, %
d\tau\,\right]\,\rho_d(t)\, %
=\, \label{e2}\\ \,=\, %
\left[\,\frac 1{m^2} %
\int_0^\infty \langle f(t,R)\, %
f(t-\tau,R-V\tau)\,\rangle\, %
(\nabla_{\!\! V} +\tau\nabla_{\!\! R})\, %
d\tau\,\right]\,\rho_d(t)\, %
\approx\, \label{e3}\\ \,\approx\, %
\left[\,\frac 1{m^2} %
\int_0^\infty \langle f(t,R)\, %
f(t-\tau,R-V\tau)\,\rangle\, %
d\tau\,\right]\,\nabla_{\!\! V}\,\rho_d(t)\, %
\,\,, \, \label{e4}
\end{eqnarray}
that is finally, in the simplest ``one-loop'' approximation, %
\begin{eqnarray}
\partial \rho_d/\partial t\,=\, %
- V\nabla_{\!\! R} \,\rho_d\,+ %
\,\nabla_{\!\! V} \,g(V)\,[(T/m) \nabla_{\!\! V} +\, %
V\,]\,\rho_d\,\, \, \label{ke} %
\end{eqnarray}

Thus, we in fact replaced  the force %
$\,f(t,R(t))\,$ by BP's coordinate-independent %
but instead velocity-dependent Gaussian white %
noise $\,\tilde{f}(t,V(t))\,$ with correlator %
\begin{eqnarray}
\langle\,\tilde{f}(t,V(t)) * \tilde{f}(t^\prime,V(t^\prime)) %
\,\rangle_V\,=\, %
2Tm\,g(V)\,\delta(t-t^\prime)\,\, %
\sim \, 2D^{(P)}\,\delta(t-t^\prime)\,
\,, \, \label{af}
\end{eqnarray}
with $\,\langle \dots \rangle_V\,$ meaning conditional %
averaging under fixed $\,V\,$. %

\subsection{Marcovian stochastic equations}

\,To write out an equivalent SE, %
notice, in view of the $\,g(V)\,$'s definition (\ref{ag}) %
and condition (\ref{c}), %
that one always can make single-valued smooth change of variables, %
$\,V\Rightarrow U\,$, such that\, %
\[
(\partial U/\partial V)\,g(V)\, %
(\partial U/\partial V)^\dagger\, =\, \overline{g} \,\,, \, \nonumber %
\]
where $\,\overline{g}\,$ is a constant (unit matrix). %
It will be good choice if it equals to $\,g(V)\,$'s %
average over equilibrium Maxwell probability %
distribution of velocity:
\begin{eqnarray}
\overline{g}\,=\,\int g(V)\, %
M(V)\, d^dV\,\,, \,\,\,\,\, %
M(V)\,=\, \frac {\exp{(-mV^2/2T)}} %
{(2\pi T/m)^{d/2}}\,\, \, \label{weq}
\end{eqnarray}
(naturally, - as Eq.\ref{ke} says, - stationary $\,V\,$'s %
distribution is Maxwellian)\,\,\footnote{\, %
Notice also that singularity of such change of variables %
would indicate inapplicability of time-local approximations %
at all, not speaking about Marcovuan approximation.}\,\,. %
Then, in terms of $\,U\,$, the SE corresponding to %
Eqs.\ref{ke}-\ref{af} looks merely as
\begin{eqnarray}
dR/dt\,=\, V(U)\,\,, \,\,\,\,\, %
dU/dt \,=\, \tilde{f}(t)/m\, - \,\gamma(U) \,\,, \, %
\label{sse}\\ 
\gamma(U)\,=\, \overline{g} \,\nabla_{\!\! U} [V^2(U) - (T/m)\,%
\ln\,\det\,g(V(U))\,]/2\,\,, \, \nonumber
\end{eqnarray}
with velocity-independent white noise source:
\begin{eqnarray}
\langle\,\tilde{f}(t) * \tilde{f}(t^\prime) %
\,\rangle_U\,=\, 2Tm\,\overline{g}\,\delta(t-t^\prime)\,\, %
\, \, \label{saf}
\end{eqnarray}

\section{Inadequacy of Marcovian approximation: \\
disclosing of conventional conjectures}

For the first look, we just demonstrated %
that our problem hides nothing novel, %
since reduces to quite trivial SE. %
But this is wrong impression. %

The matter is that the Marcovian approximation %
(\ref{ke})-(\ref{saf}) has qualitative %
defect: it neglects the above %
mentioned non-Gaussianity of the force %
acting onto BP, $\,f(t,R(t))\,$, and therefore losses %
specific non-Gaussian (higher-order) %
correlations between $\,f(t,R(t))\,$ and %
BP's path $\,R(t)\,$. %
This loss took beginning %
in transition from expression (\ref{e1}) %
to expression (\ref{e2}), %
which just means replacement of $\,f(t,R(t))\,$ by %
Gaussian white noise. %
In fact, the resulting Eqs.\ref{ke}-\ref{saf}  %
may arouse suspicions already because %
have no essential difference from equations for %
particle under short-correlated in time %
but infinitely far-correlated (constant) in space random force! %

In order to better feel importance of the loss, %
let us compare ``fidelities'' of solutions %
to SE (\ref{sse}) and LE (\ref{ale}), that is %
their sensibilities to small perturbations %
(e.g. that of initial conditions). %
At that, non-linearity of friction  %
plays no essential role, and for simplicity %
and visuality we deal with linear friction.

\subsection{Fidelity of solutions to %
SE in Marcovian approximation}

\,First, consider differential response of %
Eqs.\ref{sse}'s solutions to infinitesimally small %
change of (initial) velocity at time $\,t=0\,$, %
that is\,
\[
\,v(t)\,\equiv \,\partial V(t)/ %
\partial V(0)\,\,, \,\,\,\,\, %
r(t)\, \equiv \,\partial R(t)/ %
\partial V(0)\,\,\,\,  %
\]
Since the noise source in Eqs.\ref{sse} is %
insensible to BP's state, it %
disappears under differentiation in respect %
to $\,V(0)\,$ or $\,U(0)\,$, that is does not %
influence on $\,v(t)\,,\, r(t)\,$. %
For linear friction, when $\,U=V\,$ %
and $\,\gamma(U)=gV\,$, with %
$\,g= \overline{g} =\,$const\,, %
we thus have equations
\begin{eqnarray}
dr/dt\,=\, v\,\,, \,\,\,\,\, %
dv/dt \,=\,  - \,g\,v \,\,, \, \nonumber
\end{eqnarray}
with initial conditions $\,v(0)=1\,$, $\,r(0)=0\,$. %
Consequently, at large $\,t\,$ velocity's perturbation %
certainly tend to zero, while path's (coordinate's) one %
to a constant:
\begin{eqnarray}
v(t) \,\rightarrow\, 0 \,\,, \,\,\,\,\, %
r(t) \,\rightarrow\, 1/g \,\, \label{dse}
\end{eqnarray}

\subsection{Fidelity of solutions to %
LE in time-local friction approximation}

\,Now, turn to the approximate but more adequate %
LE (\ref{ale}). From them we have, %
also at linear friction,
\begin{eqnarray}
dr/dt\,=\,v\,\,, \, \, \,\,\, %
dv/dt \,=\, (\nabla f(t,R(t))/m)\, r \, %
- \,g\,v\,\,, \, \label{dle}
\end{eqnarray}
again wiyh initial condotions $\,v(0)=1\,$, $\,r(0)=0\,$. %
Thus, now we meet essentially multiplicative ``noise source'', %
$\,(\nabla f(t,R(t))/m)\, r\,$, which can not be made %
state-independent by a non-singular change of variabes. %
On average, solution to these equations coincides with %
(\ref{dse}). But in the sense of fluctuations it is much more %
interesting: naturally, it is statistically unstable. %

To see this in most simple way, let us consider 1D case, %
$\,d=1\,$, - when $\,v(t)\,$, $r(t)\,$, etc., become scalars %
instead matrices. Introduce random DF
\[
\widetilde{\varrho}(t,r,v)\,=\, %
\delta(r-r(t))\, \delta(v-v(t))\,\, \,
\]
and derive approximate kinetic equation for its %
average
\[
\varrho(t,r,v)\,=\, \langle\, %
\widetilde{\varrho}(t,r,v)\,\rangle\,\,, \, %
\]
by treating $\,f(t,R(t))\,$ as white noise %
(like in (\ref{e2})-(\ref{e4})). %
The result is
\begin{eqnarray}
\partial \varrho/\partial t\,=\, %
[\,- v\nabla_{\!\! r} \, +\, %
g\nabla_{\!\! v} \,v\,]\, \varrho\,+ %
\,Q\,(r\nabla_{\!\! v})^2\, \varrho\,\,, \, \label{dke} %
\end{eqnarray}
where parameters are expressed by
\begin{eqnarray}
g\,=\,\overline{g}\,=\, %
\frac 1{2mT} \int k^2\, %
\overline{S}(|k|)\, dk\,\,, \, %
\,\,\,\, %
Q\,\equiv\,  %
\frac 1{2m^2} \int k^4\,\overline{S}(|k|)\, %
dk\,\,, \, \label{p}\\
\overline{S}(|k|)\,\equiv\, %
\int S(Vk,|k|)\, M(V)\, dV\,\,\, \nonumber
\end{eqnarray}

Next, considering evolution of second-order %
statistical moments, from Eq.\ref{dke} (or directly %
from Eqs.\ref{dle} we have
\begin{eqnarray}
\frac d{dt}\, %
\left(%
\begin{array}{c}
  \langle r^2\rangle \\
  \langle rv\rangle \\
  \langle v^2\rangle \\
\end{array}%
\right)
\,=\, \mathcal{M}\, %
\left(%
\begin{array}{c}
  \langle r^2\rangle \\
  \langle rv\rangle \\
  \langle v^2\rangle \\
\end{array}%
\right)
\,\,\,, \,\,\,\,\, %
\mathcal{M}\,=\, %
\left(%
\begin{array}{ccc}
  0 & 2 & 0 \\
  0 & -g & 1 \\
  2Q & 0 & -2g \\
\end{array}%
\right) \,\,\,  \label{sm}
\end{eqnarray}
Eigenvalues, $\,\mu_j\,$, of matrix %
$\,\mathcal{M}\,$ are roots of cubic equation
\begin{eqnarray}
\mu\,(\mu +g)\,(\mu + 2g)\,=\, 4Q \,\, \label{root}
\end{eqnarray}
It clealy shows that one of roots, - %
let be denoted by $\,\mu_+\,$, - %
 is real positive, %
that is solutions of Eqs.\ref{dle} are unstable %
in the sense of second-order (and hence higher-orfer) moments. %

Rate of the instability,\, $\,\mu_+\,$, %
as compared with velocity %
relaxation rate $\,g\sim 1/\tau_v\,$, is determined by dimensionless %
paramer $\,Q/g^3\,$. %
From (\ref{p}) it follows that
\begin{eqnarray}
\frac Q{g^3}\,\sim\, \frac %
T{m\,r_c ^2 g^2}\,\sim\, %
\frac {\lambda^2}{r_c ^2}\,\,, \, \nonumber
\end{eqnarray}
where\, $\,r_c \,$\, is characteristic correlation %
length of the force field $\,f(t,r)\,$, and %
$\,\lambda \sim \sqrt{T/m}/g \sim %
\tau_v \sqrt{T/m}\,$\, characteristic BP's %
``free path'' length. %
Hence, if $\,r_c \ll \lambda\,$, %
then $\,Q/g^3 \gg 1\,$ and, according to Eq.\ref{root}, %
$\,\mu_+ \gg g\,$.   %

\subsection{Fundamental incompleteness of Marcovian approach %
and typical ``bottleneck'' of dynamical theory of noises}

\,We just revealed crucial defect of %
Marcovian approximation: it completely losses %
exponential instability of BP's trajectories in respect to %
their small perturbations. %
As the consequence, it losses all statistical effects of %
this instability and therefore, generally speaking, %
can give only a caricature of real dynamical noise. %

Then, how one should avoid the loss? The answer %
was prompted long ago by critical analysis %
of ``molecular chaos'' in %
fluids \cite{i1}, crystals \cite{i3}, %
under charge transport \cite{ufn,pr157,bk12} %
and generally in transport phenomena \cite{i2,pr157}. %
Namely, we have to reject any {\it \,a priori\,} %
statements (even very attractive) about %
``independencies'' of ``random'' %
variables, - like e.g. ``molecular chaos hypothesis'' %
or ``marcovianity'', - and allow any statistical %
dependencies and correlations compatible with %
equations of statistical mechanics. %
After that we might find that some of {\it \,a priori\,}  %
unexpected or neglected dependencies and correlations %
really take place and are physically important. %

Thus, in general there are two variants of approximate %
theory of transport noise: one (conventional) before %
the mentioned theoretical ``bottleneck'' %
(overcoming usual instinctive %
conjectures\,\footnote{\, %
``Prejudices'' disclosed by N.\,Krylov in \cite{kr}. %
}\,) %
and another behind the ``bottleneck''. %
Figuratively speaking, that are two different solutions %
of same approximate equations, ``trivial'' and %
``non-trivial''. The latter contains low-frequency %
fluctuations, - like 1/f-noise, - of transport characteristics %
which are constants in the former\,\footnote{\, %
For more explanations see review-discussion pages %
in \cite{j,o,ufn,i1,i2,i3,p157,bk12} and %
also \cite{p1,0710,p0802,p0803,p0806,tmf, %
p1105,p1203,p1209,p1207}. A hidden role of the exponential %
instability in fundamental 1/f-noise formation was %
directly demonstrated \cite{i3}.}\,. %


\section{First steps to solution of the problem. %
Exponential instability, higher-rder statistics, %
and scaleless diffusivity fluctuations}

\,To understand possible consequences %
of the exponential instability, %
first it is useful to point out several facts.

\subsection{Some useful formulae and remarks}

\,\,\,

{\bf 1}.\, Stationary %
(equilibrium) distribution of BP's velocity has %
Maxwellian form (\ref{weq}), i.e. is Gaussian, regardless of degree of %
BP-thermostat interaction. %
This statement follows from the structure  of %
Hamiltonian of our system as defined by %
(\ref{h}) and (\ref{hh}). %
Therefore, in equilibrium statistical ensemble, %
for any function $ \,\Phi(V,\dots)\,$ %
of $\,V\,$ and some other random factors %
independent on $\,V\,$, we can write
\begin{eqnarray}
\langle\,V,\,\Phi(V,\dots)\,\rangle\, = %
(T/m)\,\langle\,\nabla_{\!\! V}  \Phi(V,\dots)\,\rangle\,
 \,, \, \,\,\,\,  \label{gv}\\ %
\nonumber %
\langle\,V,V,\,\Phi(V,\dots)\,\rangle\, = %
(T/m)^2\,\langle\,\nabla_{\!\! V} %
\nabla_{\!\! V}\, \Phi(V,\dots)\,\rangle\, \,, \,
\end{eqnarray}
etc. Here and below, angle brackets with $\,n\,$ commas denote %
joint cumulant of $\,n+1\,$ expressions separated by the commas %
(the Malakhov's cumulant brackets \cite{mal}). %
Similarly, since field $\,f(t,r)\,$ is Gaussian,
\begin{eqnarray}
\langle\,f(t,r),\,\Phi(V,f,\dots)\,\rangle\, = %
\int dt^\prime \int dr^\prime\, %
\langle\,f(t,r)\,,\,f(t^\prime,r^\prime)\,\rangle\, %
\left\langle\,\frac {\delta \Phi(V,f,\dots)}{\delta %
f(t^\prime,r^\prime)}\,\right\rangle\, %
\, \, \,  \label{gvf} %
\end{eqnarray}
(so-called Furutsu-Novikov formula).

{\bf 2}.\, BP's state $\,\{R(t_0),V(t_0)\}\,$ %
at arbitrary chosen (and then fixed) %
``initial'' time moment $\,t_0\,$ %
can be considered as statistically independent %
on the simultaneous medium's state. %
Then, any function of later BP's states $\,\{R(t),V(t)\}\,$ %
($\,t>t_0\,$) and BP's displacement (path) %
$\,\Delta R=R(t)-R(t_0)\,$ %
gives an example of the mentioned function %
$ \,\Phi(V,\dots)\,$, with $\,V=V(t_0)\,$, %
and Eq.\ref{gv} yields, in particular,
\begin{eqnarray}
\langle\,V(t),V(t_0)\,\rangle\, = %
(T/m)\,\langle\,\partial V(t)/\partial V(t_0)\,\rangle\,
 \,, \, \,\nonumber\\ %
\langle\,\Delta R,V_0\,\rangle\, = %
(T/m)\,\langle\,\partial \Delta R/\partial V_0\,\rangle\,
 \,, \, \,   \label{gv1}\\
\langle\,\Delta R^2,V_0,V_0\,\rangle\, = %
(T/m)^2 \langle\,\partial^2 \Delta R^2/ %
\partial V^2_0\,\rangle\, \,, \, \,   \label{gv2}
\end{eqnarray}
where for brevity we use $\,V_0\equiv V(t_0)\,$.

{\bf 3}.\, Notice that at $\,t-t_0\gg \tau_v\,$ %
second of expressions (\ref{gv1}) %
gives BP's diffusivity, let be denoted by $\,D\,$. %
According to Eqs.\ref{dse}, %
$\,D\approx T/mg =T/m\overline{g}\,$. %
Thus, mean value of the differential response %
$\,\partial \Delta R/\partial V_0\,$ is directly %
connected to the diffusivity.

Similarly, fluctuations of this response are %
closely connected to BP's diffusivity fluctuations. %
The latter, on the other hand, can be adequately characterized %
by fourth-order BP's path-velocity cumulants. %
Namely \cite{o,ufn,i1,i2,pr157,bk12,p1,p0802,p1105,p1209,p1207}, %
at $\,\tau \equiv t-t_0 \gg \tau_v\,$, function
\begin{eqnarray}
K_D(\tau)\,=\, \frac 1{24} %
\frac {d^2}{d\tau^2}\, %
\langle\,\Delta R,\Delta R,\Delta R,\Delta R\,\rangle\, %
\,=\, \nonumber\\ =\, %
\frac 1{24} \frac {d^2}{dt^2}\, %
\langle\,\Delta R^{(4)}\,\rangle\, %
\,=\, \frac 1{24} \frac {d^2}{dt_0^2}\, %
\langle\,\Delta R^{(4)}\,\rangle\, %
\,=\, -\,\frac 1{24} \frac {d^2}{dt\,dt_0}\, %
\langle\,\Delta R^{(4)}\,\rangle\, %
\, \, \, \,   \label{kd}
\end{eqnarray}
plays role of correlation function of equilibrium %
BP's diffusivity fluctuations (and that of BP's %
mobility fluctuations in weakly non-equilibrium regime %
under external force). %
Here $\,\langle\,X^{(n)}\,\rangle\,$ is short %
designation for $\,X\,$'s $\,n\,$-order cumulant, %
and we took into account that at %
$\, t-t_0 \gtrsim \tau_v\,$ the path statistics must %
depend on time difference $\,t-t_0\,$ only. %
We thus have
\begin{eqnarray}
K_D(t-t_0)\,=\, \frac 1{2} %
\, \langle\,V(t)\,,\,\Delta R, \Delta R\,,\, %
V(t_0)\,\rangle\, \,, \, \,   \label{kd1}
\end{eqnarray}
or equivalently
\begin{eqnarray}
K_D(t-t_0)\,=\, \frac 1{2} %
\, \langle\,\Delta R, \Delta R, V(t_0), %
V(t_0)\,\rangle\,-\, %
\frac 1{6} %
\, \langle\,\Delta R, \Delta R, %
\Delta R, \,dV(t_0)/dt_0\,\rangle\, %
\,=\, \label{kd2}\\ \,=\, %
\, \frac 1{2} %
\, \langle\,V(t), V(t)\,,\, %
\Delta R, \Delta R\,\rangle\,+\, %
\frac 1{6} %
\, \langle\,dV(t)/dt\,,\,  \Delta R, \Delta R, %
\Delta R \,\rangle\, \,\,\nonumber %
\end{eqnarray}

The differential response appears in visual form  %
if we notice that at $\,t-t_0\gg \tau_v\,$ %
the two velocity values, $\,V(t)\,$ and $\,V(t_0)\,$, %
certainly are almost statistically independent and hence %
mutually Gaussian random quantities. Consequently, %
formula (\ref{kd1}) can be transformed as follows, %
\begin{eqnarray}
K_D(t-t_0)\,\Rightarrow\,  %
\frac {T^2}{m^2}\,\left[\, \frac 12\, \left\langle\, %
\frac {\partial^2 \Delta R^2} %
{\partial V(t) \,\partial V(t_0)}\, %
\right\rangle\,-\, %
\left\langle\, \frac {\partial \Delta R} %
{\partial V(t)}\,\right\rangle\, %
\left\langle\, \frac {\partial \Delta R} %
{\partial V(t_0)}\,\right\rangle\, %
\right]\, \,, \, \,   \label{kd1_}
\end{eqnarray}
Clearly, right-hand side here consists of %
squared first-order differential response and besides %
second-order (double-differential) one.
Analogously, with the help of Eqs.\ref{gv}-\ref{gvf} %
and other above formulas, transforms expressions (\ref{kd2}).

{\bf 4}.\, Integrating Eq.\ref{kd}, one in the same fashion %
obtains relations
\begin{eqnarray}
\int_0^\tau K_D(\tau^\prime)\,d\tau^\prime \,=\,  %
\frac 16\,\left\langle\, %
\Delta R ,\Delta R, \Delta R ,\, V_0\, \right\rangle\,
\,, \, \nonumber\\
\frac 1\tau \int_0^\tau K_D(\tau^\prime)\,d\tau^\prime \,=\,  %
\frac 1\tau\, \frac {T}{2m}\,\left\langle\, %
\Delta R ,\Delta R,\, \frac {\partial \Delta R} %
{\partial V_0}\, \right\rangle\,\, \, \label{kd3}
\end{eqnarray}
The latter thus calls for analysis of third-order  %
irreducible correlations (cumulants) which address %
$\,K_D(\tau\,$ to mutual correlation between %
 ``fidelity'' and ``diffusivity'' of BP's trajectories. %
In essence, this is particular case of general exact expression %
for correlation functions of low-signal excess noise and %
dissipation fluctuations \cite{i3} (see also \cite{bkn}).


\subsection{How the instability might work}

\,\,\,

{\bf 1}.\, In the Marcovian approximation, %
the fourth cumulant $\,\langle \Delta R^{(4)} %
\rangle\,$ certainly is either linear function of %
$\,\tau =t-t_0\,$ at $\,\tau \gtrsim \tau_v\,$ or %
(in linear approximation) identical zero. %
Hence, $\,K_D(\tau)\,$ can appear non-zero at %
$\,\tau \gg \tau_v\,$ due only to what is lost under %
Marcovian approximation, i.e. the exponential instability of %
BP's trajectories. %
Hence, to calculate some of the $\,K_D(\tau)\,$'s %
expressions (\ref{kd})-(\ref{kd3}), one have to %
substitute there exact solutions of LE (\ref{le}) %
or (\ref{le2}) and then perform necessary averaging %
over realizations of the field $\,f(t,r)\,$. %
However, it is impossible in literal sense, %
just because of the exponential instability. %

At present, we do not know a regular method %
to break away from this ``vicious circle'' %
(penetrate through the ``bottleneck''). %
Therefore,  it would be quite good if we  %
demonstrated significance of $\,\langle \Delta R^{(4)} %
\rangle\,$ and $\,K_D(\tau)\,$ at $\,\tau \gg \tau_v\,$ %
at least under some reasonable approximation of LE. %

In this respect, the last of equivalent expressions %
(\ref{kd})-(\ref{kd3}) appears most suitable, in combination %
with the approximate time-local Eqs.\ref{ale} and \ref{dle}, %
since this combination most visually highlights statistical interference %
between BP's trajectory $\,\{R(t),V(t)\}\,$ itself and its %
fidelity $\,\{r(t),v(t)\}\,$.

{\bf 2}.\, For simplicity, moreover, we apply also linear-friction %
approximation, - replacing in Eqs.\ref{ale} %
$\,g(V)\,$ by $\,g=\overline{g}=\,$const\, (with %
$\,\overline{g}\,$ from (\ref{weq})), - and, besides, %
calculate right-hand side in Eq.\ref{kd3} under %
$\,V_0=V(t_0)=0\,$ (for anyway long-time behavior of %
$\,K_D(\tau)\,$ must be indifferent %
to $\,V(t_0)\,$\,\, \footnote{\, %
At that, we merely omit from solution $\,\Delta R\,$ %
of Eqs.\ref{ale} additive $\,V_0\,$'s contribution, %
$\,\exp{(-gt)}\,V_0\,$.}\,). %

Then, introducing functions
\[
\xi(t)\,= \,f(t,R(t))/m\,\,, \,\,\,\, %
\eta(t)\,= \, \nabla f(t,R(t))/m\,\,, \,\,\,\, %
C(\tau)\,=\,[\,1\,-\,\exp{(-g\tau)}\,]/g\,\,, \,
\]
and taking $\,t_0=0\,$ and naturally $\,R(0)=0\,$, %
we can write solution to Eqs.\ref{ale} as
\begin{eqnarray}
\Delta R(t)\,= \int^t_0 C(t-t^\prime)\, %
\xi(t^\prime)\, dt^\prime\,\,,\,  \label{rt}
 \end{eqnarray}
while solution to Eqs.\ref{dle} as infinite iteration series
\begin{eqnarray}
r(t)\,=\,C(t)\,+ \int^t_0 C(t-t^\prime)\, %
\eta(t^\prime)\,r(t^\prime)\, dt^\prime\,=\, \nonumber\\
=\,C(t)\,+ \int^t_0 C(t-t^\prime)\, %
\eta(t^\prime)\, C(t^\prime)\,dt^\prime\,+ %
\, \label{drt}\\ +\, %
\int^t_0 C(t-t^\prime)\, %
\eta(t^\prime)\, \int_0^{t^\prime} %
C(t^\prime-t^{\prime\prime})\, %
\eta(t^{\prime\prime})\, C(t^{\prime\prime})\, %
dt^\prime\,dt^{\prime\prime}\,+\, \dots\,\, \nonumber
\end{eqnarray}
Next, inserting all this into Eq.\ref{kd3} %
and imagining, - in the spirit of time-local %
linear approximation, - $\,\xi(t),\,\eta(t)\,$ %
like white noises, one can see that %
in fact the only third term of expansion (\ref{drt}) %
survives after averaging. %
It produces %
\begin{eqnarray}
\int_0^t K_D(\tau)\,d\tau \,=\,  %
\frac {T}{m}\,\langle\, %
\Delta R(t) ,\Delta R(t),\,r(t)\,\rangle\,\approx\,   \, %
\label{akd}\\ \approx\, %
\frac {T}{m} \int^t_0 dt^\prime %
\int_0^{t^\prime} dt^{\prime\prime}\, %
C^2(t-t^\prime))\, C(t^\prime-t^{\prime\prime})\, %
C(t-t^{\prime\prime})\, C(t^{\prime\prime})\, %
\,\times\, \nonumber\\ \times\, %
\int \!\!\int %
\langle\,I(t_1-t^\prime,R(t_1)-R(t^\prime))\, %
I(t_2-t^{\prime\prime},R(t_2)-R(t^{\prime\prime}))\, %
\rangle\, dt_1\,dt_2\, \,\, \nonumber
\end{eqnarray}
Here new (tensor) function $\,I(\tau,r)\,$ %
appears defined by  ``$\eta$-$\xi$'' cross-correlator:
\begin{eqnarray}
\langle\, \nabla f(t^\prime,R(t^\prime)) * %
f(t_1,R(t_1))\,\rangle_R/m^2\, %
\equiv \,I(t_1-t^\prime, R(t_1)-R(t^\prime))\,\,, \,\nonumber\\
I(\tau,r)\,=\, \nabla*\nabla*\nabla\, K_{xx}(\tau,r)/m^2\,=\, %
\label{i}\\ \,=\, %
\frac 1{m^2}\int\!\!\int k*k*k\,\, %
\sin\,kr\,\, \cos\,\omega\tau\,\, %
S(\omega,|k|)\, \frac {d\omega}{2\pi}\, %
dk\,\,, \, \nonumber
\end{eqnarray}
with $\,\langle\dots \rangle_R\,$  %
standing for conditional average under given %
BP's trajectory. %

Evidently (and importantly), %
when performing integrations over $\,t_1\,$ %
and $\,t_2\,$ in Eq.\ref{akd}, we have to make replacement
\begin{eqnarray}
R(t_1)-R(t^\prime)\,\Rightarrow\, %
V(t^\prime)\,\tau\,-\,g\,V(t^\prime)\, %
\tau^2/2\, \,\, %
\,\,\,\,\, (\,\tau \equiv t_1-t^\prime\,)\,\,, %
\, \label{rep}
\end{eqnarray}
and similarly for $\,R(t_2)-R(t^{\prime\prime})\,$. %
At that, we remove from this displacements %
their parts containing $\,f(t,R(t))\,$, %
since contribution of these parts to  the average in (\ref{akd}) %
is definitely negligible, at least under %
condition $\,\tau_v\gg \tau_c\,$. %
Accordingly, - applying under integral in Eq.\ref{i} approximation %
\[
\sin{[k\,(V\tau -gV\tau^2/2)]}\, %
\approx\,\sin\, kV\tau\,\, -\,k\,(gV\tau^2/2)\, %
\cos\, kV\tau\,\,, \,
\]
- we can write
\begin{eqnarray}
\int I(t_1-t^\prime,R(t_1)-R(t^\prime))\,dt_1\, %
\Rightarrow\, \,\,g\,\Upsilon(V(t^\prime))\,%
V(t^\prime)\,\,, \, \label{rep1}\\ %
\Upsilon(V)\,\equiv \, %
\frac 1{m^2} \int k^4\, %
S^{\prime\prime}(kV,|k|)\,dk\,=\, %
\frac {2T}m \, \nabla_{\! V}^2\,g(V)\, %
\, \, \label{ups}\\
\left(\,S^{\prime\prime}(\omega,|k|)\,\equiv\, %
\frac {\partial^2 S(\omega,|k|)}{\partial \omega^2}\, %
\right) \, \,, \, \nonumber
\end{eqnarray}
and analogously for second multiplier under the average %
(using symbolical scalar notations, instead of tensor %
ones, or for simplicity taking in mind $\,d=1\,$).

At last, before inserting (\ref{rep1}) to (\ref{akd}) %
let us make simplification as follows, %
\begin{eqnarray}
\langle\,g\,\Upsilon(V(t^\prime))\,%
V(t^\prime)\, g\,\Upsilon(V(t^{\prime\prime}))\,%
V(t^{\prime\prime})\,\rangle\, \Rightarrow\, %
g^2\overline{\Upsilon}^2\, %
\langle\,V(t^\prime)\,V(t^{\prime\prime})\,\rangle\, %
\approx\, \label{rep2}\\ \approx\, %
g^2\overline{\Upsilon}^2\, %
\frac Tm\, \exp{(\,-g\,|t^\prime - %
t^{\prime\prime}|)}\,\, \, \nonumber\\
\left(\, %
\overline{\Upsilon}\,\equiv\, %
\int \Upsilon(V)\,M(V)\,dV\, %
\right)\,\,, \, \nonumber
\end{eqnarray}
and besides notice that at $\,t\gg \tau_v\sim 1/g\,$ %
all functions $\,C(\dots)\,$ in Eq.\ref{akd}), - %
except $\,C(t^\prime -t^{\prime\prime})\,$, - %
can be replaced by constant $\,1/g\,$\, %
\footnote{\, Notice also that in place of %
$\,\overline{\Upsilon}^2\,$ it may be more correct to write %
$\,\overline{\Upsilon^2}=$ $\int \Upsilon^2(V)M(V)\,dV\,$. %
}. %
Then we come to
\begin{eqnarray}
\frac 1t \int_0^t K_D(\tau)\,d\tau \,\approx\,  %
\frac {T^2\,\overline{\Upsilon}^2}{2\,m^2\,g^4}\, %
=\, D^2\,\frac {\overline{\Upsilon}^2}{2\,g^2}\, %
\,, \, \label{aakd}
\end{eqnarray}
that is to non-decaying, infinitely long-range, %
diffusivity's correlation function. %


{\bf 3}.\, Let us recall that, by the $\,K_D(\tau)\,$'s %
``microscopic'' definition (\ref{kd}) \cite{o,pr157,bk12,i1,i2,p1,p1207}, %
in ``macroscopic'' (phenomenological) sense
\[
K_D(\tau)\,=\,\langle\,\widetilde{D}(t+\tau)\,,\, %
\widetilde{D}(t)\,\rangle\,=\, %
\langle \widetilde{D}(t+\tau)\, %
\widetilde{D}(t)\rangle\,-\, %
\langle\widetilde{D}\rangle^2\,\,, \,
\]
where $\,\widetilde{D}(t)\,$ represents fluctuating diffusivity, %
and $\,\langle\widetilde{D}(t)\rangle=D\,$. %
Hence, our result (\ref{aakd}) states that correlation function %
of BP's diffusivity fluctuations never decays to zero, %
as if $\,\widetilde{D}(t)\,$ were constant in time but %
randomly different from one BP's trajectory to another. %

Such ``quasi-static'' fluctuations are typical %
result of quantitatively rough theoretical approaches to %
1/f-type low-frequency fluctuations of diffusivity/mobility %
or other transport rates. For examples see %
e.g. \cite{kmg,p1007}. %
Nevertheless, nothing prevents such approaches from giving %
reasonable estimates of 1/f-noise level, and they are rather %
correct in prediction of characteristic long-range statistical scale %
invariance of transport processes \cite{ufn,pr157,bk12,i1,i2,i3}, %
i.e. $\,\Delta R=R(t)-R(0)\,$ in the present case. %

Indeed, considering higher-order equilibrium cumulants %
$\,\langle \Delta R^{(2n)}\rangle\,$ by means of obvious %
generalization of exact relation (\ref{kd3}),
\begin{eqnarray}
\frac d{dt}\, \langle\, %
\Delta R^{(2n)}\,\rangle\,=\, 2n\,(2n-1)\, %
\frac Tm\, \left\langle\, %
\Delta R^{(2n-2)}\,,\, \frac {\partial \Delta R} %
{\partial V_0}\, \right\rangle\,\,, \, \label{hm}
\end{eqnarray}
and again approximate expressions (\ref{rt})-(\ref{drt}), %
it is not too hard to see that (at $\,t\gg \tau_v\,$)
\begin{eqnarray}
\langle\, \Delta R^{(2n)}\,\rangle\,\approx\, %
(2n-1)!!\,c_n\, \langle \Delta R, \Delta R \rangle^n\, %
\propto\, t^n\,\,, \,  \label{hmas}
\end{eqnarray}
with some coefficients $\,c_n\,$ which can be obtainedd %
from a recursive procedure. %
Such the asymptotic law detects essential non-Gaussianity of %
transport process and inapplicability of the ``law of large %
numbers'' to it.

At the same time, more accurate theories %
expectedly must lead to violation of such literal scale %
invariance and appearance of some slow-varying, %
logarithmic or power-law, factors in the $\,K_D(t)\,$ %
and higher-order cumulants (\ref{hmas}). For examples see %
e.g. \cite{ufn,pr157,bk12,i1,i2,p1,p0802,p0806,tmf,p1209,p1007}.

\section{Variance of diffusuvity fluctuations. %
Discussion of the result} %

{\bf 1}.\, Testing of more correct self-consistent %
approaches to our present problem we leave for future. %
Now, instead let us discuuss our spare %
but not trivial result (\ref{aakd}). %

Its above derivation %
shows that formal ``entry point'' to the mentioned theoretical %
``narrow bottleneck'' may be accounting for %
interplay between the exponential instability %
and friction (dissipation) both simultaneously induced by %
the medium (thermostat). Indeed, non-zero value of the %
integrated ``$\,\eta$-$\xi\,$'' correlation in Eq.\ref{i} %
is due to the second term of (\ref{rep}) %
reflecting BP's ``braking'' by the friction.   %
Hence, in essence that is ``instability-friction'' correlation. %
It then naturally causes fluctuations %
in friction-related characteristics of BP's %
motion, first of all in diffusivity, as Eq.\ref{kd3} promps. %

At that, the complete  ``$\,I$-$I\,$'' correlor %
in Eq.\ref{akd} (and its simplification in Eq.\ref{rep2}) is in fact %
fourth-order cross-correlator (cumulant) of the random %
force $\,f(t,R(t))\,$ and its gradient $\,\nabla f(t,R(t))\,$. %
According to Eqs.\ref{kd3}, \ref{akd} and finally \ref{aakd}, %
this correlation implies specific long-range %
fourth-order (four-point) irreducible BP's velocity correlations. %
They, in turn, in accordance with Eq.\ref{hm}, %
give rise to an infinite hierarchy of higher-order %
(many-point) long-range velocity (and force) cumulants. %

{\bf 2}.\, Physical meaning of all such terrible picture %
was not once commented and explained in our works during last %
thirty years (please see above references and that therein).  %

Our consideration once again demonstrated that %
diffusivity by its origin never can be quite certain ({\it \,a priori\,} %
predictable) quantity. In absence of friction, it %
would have no definite value at all, because of constant BP's %
stochastic acceleration and kinetic energy growth %
(as if the medium had infinitely %
large temperature). But when friction takes place, %
it immediately interferes with the exponential %
instability (IE) and, - together with BP's diffusivity, - %
 acquires fluctuations,  as unboundedly %
diverse and unique as IE is in itself %
in (infinitely) many-particle systems. %

In principle, such consequences of IE were predicted %
already by N.\,Krylov \cite{kr}.

{\bf 3}.\, What is for quantitative meaning of our result %
(\ref{aakd}), there are many different possibilities %
depending on structure of the medium's spectral function %
$\,S(\omega,|k|)\,$. Therefore let us imagine that it %
possesses some primitive ``bell-like'' shape and can be %
characterized by three parameters only, i.e. medium's %
correlation time $\,\tau_c\,$ and length $\,r_c\,$ and, %
besides, magnitude of the force field fluctuations. %
The latter parameter can be replaced  %
by the velocity relaxation rate $\,g\sim 1/\tau_v\,$.

Then, in view of Eq.\ref{ups} and under approximation %
(\ref{rep2}), we come to estimate %
\begin{eqnarray}
\frac {K_D}{D^2} \,\approx\,  %
\frac {\overline{\Upsilon}^2}{2g^2}\, %
\sim\, \left(\frac Tm\, \frac {\tau_c^2}{r_c^2} %
\right)^2\,\sim\, \left(\frac T{m\,u^2} \right)^2\, %
\,, \, \label{est}
\end{eqnarray}
where\, $\,u\sim r_c/\tau_c\,$ can be interpreted %
as characteristic propagation velocity of the force field. %
Thus, this estimate well agrees with the reasonings %
expounded in Sec.7 and 9. %

Let us also remind of, firstly, our assumption $\,g\ll 1/\tau_c\,$, %
which in standard terms means weakness of the field %
fluctuations. Secondly, conventional opinion that %
Marcovian approximation coincides with exact theory at least %
in the ``infinitely weak interaction (weak noise) limit'' %
$\,g\tau_c\rightarrow 0\,$. %
In reality, however, - as Eq.\ref{est} shows, - %
variance of relative low-frequency (quasi-static) %
 diffusivity fluctuations, $\,\widetilde{D}(t)/D\,$, %
is  insensible to the $\,g\tau_c\,$'s value, if the %
ratio $\,\tau_c/r_c\,$ keeps constant. %
Hence, generally Marcovian approximation has no %
justification even in the weak noise limit (not speaking %
that anyway it losses any diffusivity %
fluctuations)!\, \footnote{\, %
Unfortunately, conventional ideal %
of randomness and ``stochasticity'' appeals to %
white noise (or trivially related ``colored noises'') %
but presumes no place for something like our time-scaleless diffusivity's %
(or other kinetic rates') fluctuations, even if %
being sowed to dynamical ground. Such an example %
is given by \cite{acc}. It shows that %
a biassed ideology may emasculate even mathematical physics. %
All the more so since in practice growth of %
scientific knowledge is accompanied %
by that of scientific prejudices and fallacies \cite{zin}, %
which was excellently illustrated in \cite{kr} by %
example of statistical physics. %
In contrast to poor pure stochastjcs, dynamics %
allows for some ``free will''.
}\, %

To improve our estimates of the diffusivity 1/f noise, %
even in case weak medium's noise, one should return to formally %
exact LE (\ref{le}) or (\ref{le2}) %
or may be (\ref{hj_}) and carefully do with their non-linearities, %
playing significant role at non-zero ratio %
\,$\,\tau_c/r_c\,$\,.


\section{Conclusion}

To resume, we considered Langevim equations describing random walk of %
particle in thermodynamically equilibrium fluctuating medium, %
and showed that the particle's diffusivity undergoes %
scaleless (1/f-type) low-frequency fluctuations whose %
magnitude can be comparable with average value of diffusivity %
(or even much exceed it) regardless of magnitude of the %
medium noise.

At that we demonstrated, on one hand, usefulness of %
traditional ``stochastic calculus'' in the framework of %
dynamically based theory. On the other hand, %
necessity to control stochastic way of thinking, since its %
seeming completeness may lead to too hasty and wrong %
conclusions (in particular it by itself automatically losses %
the diffusivity fluctuations under our interest).

The obtained result confirms rather %
general ``theorem on fundamental 1/f noise'' discussed in %
\cite{p1207} (and, under more specific conditions, %
in \cite{o,p1105}). I hope it will stimulate further %
investigations of transport 1/f noises in various %
many-particle Hamiltonian systems.

\begin{eqnarray}
\end{eqnarray}

\,\,\, 

\end{document}